\documentclass{nature_fig}

\usepackage[english]{babel}
\usepackage{bbold}
\usepackage{bbm}
\usepackage[pdftex]{graphicx}
\usepackage{latexsym,amsmath,verbatim}
\usepackage{color}
\usepackage{rotating}
\usepackage{multirow}
\usepackage{color}
\usepackage{caption}

\usepackage{amssymb,amsfonts,amsmath}

\newcommand{\Eq}[1]{Eq.\,(\ref{#1})}

\renewcommand{\eqref}{\Eq}

\usepackage{comment}

\newcommand{\rev}[1]{{\color{black}#1}}

\newcommand{\heading}[1]{{\vspace{0.25truecm}\noindent\textbf{#1.}}}

\usepackage{xcolor}
\definecolor{RoyalBlue}{HTML}{4169e1}
\definecolor{ForestGreen}{HTML}{228b22}

\usepackage[colorlinks = true,
            linkcolor = blue,
            urlcolor  = blue,
            citecolor = blue,
            anchorcolor = blue]{hyperref}

\usepackage{comment}
\usepackage{subcaption}

\usepackage{todonotes}

\title{Quantifying efficient information exchange in real network flows}

\author{Giulia Bertagnolli$^{1,2}$, Riccardo Gallotti$^{1}$, Manlio De Domenico$^{1\ast}$}

\begin{document}

\maketitle

\begin{affiliations}
\item CoMuNe Lab, Fondazione Bruno Kessler, Via Sommarive 18, 38123 Povo (TN), Italy
\item University of Trento, Department of Mathematica, Via Sommarive, 14, 38123 Povo (TN), Italy
\end{affiliations}
\noindent Corresponding author: mdedomenico@fbk.eu
\date{\today}

\baselineskip24pt


\begin{abstract}
Network science enables the effective analysis of real interconnected systems, characterized by a complex interplay between topology and interconnections strength. It is well-known that the topology of a network affects its resilience to failures or attacks, as well as its functions.
Exchanging information is crucial for many real systems: the internet, transportation networks and the brain are key examples.
Despite the introduction of measures of efficiency to analyze network flows, i.e. topologies characterized by weighted connectivity, here we show that they fail to capture combined information of link existence and link weight.
Here we propose a physically-grounded estimator of flow efficiency which can be computed for every weighted network, regardless from the scale and nature of weights and from any (missing) metadata.
Remarkably, results show that our estimator captures the heterogeneity of flows along with topological differences and its complement information obtained from percolation analysis of several empirical systems, including transportation, trade, migrations, and brain networks. \rev{Additionally,} cutting the heaviest connections may increase the average efficiency of the system and hence, counterintuively, a sparser network is not necessarily less efficient. Our estimator enables the comparison of communication efficiency of disparate systems, without the possible pitfalls deriving from the scale of flow.
\end{abstract}


\section*{Introduction}

Complex systems store energy, process and, very often, efficiently exchange information to perform complex tasks. The universal mechanisms behind this behavior are unknown, although pioneering works have shown that the robustness of this type of systems to random failures or targeted attacks~\cite{albert2000error} might emerge from the trade-off between the cost of exchanging information and the importance of guaranteeing communication dynamics for functioning~\cite{watts1998collective,latora2001efficient,avena2018communication}.
Therefore, it is crucial for units in a complex network to route information through shortest paths, broadcasting or according to some dynamics between these two extremes~\cite{yan2006efficient}, as it happens for instance in the Internet~\cite{huitema2000routing}.
For several applications of interest, even the inverse problem, of identifying either the origin or the destination of the flow from the observation of pathways, is relevant~\cite{goltsev2012localization,pinto2012locating}.
This framework enables the description of a wide variety of systems, from cell signaling to individuals exchanging information in social/socio-technical systems such as human flows through different parts of a city by public or private transportation means.
In the following we focus our attention on flow networks, systems characterized by the exchange of flows -- e.g., number of streets between different parts of the city or human movements within a city, migration between different geographic areas, goods traded among countries, packets routed among servers, electricity in a power grid  -- through edges \cite{lima2015disease,rosvall2014memory,Li669,arianos2009power}.
System's units and their connections have a limited capacity and, in absence of sources and sinks, the sum of the overall incoming and outgoing flows is constant. One widely accepted measure of efficiency in information flow is the global communication efficiency, that has been used to highlight the possible designing principles responsible for neural, man-made communication, and transportation systems~\cite{latora2001efficient}.

Here we show that a normalized descriptor of global efficiency can be computed without any knowledge on the system, but its weighted network representation. In fact, for a wide class of weighted systems~ \cite{barrat2004architecture} which are not embedded in space or for which metadata about the underlying geometry (nodes coordinates) are not available, the classical global efficiency might be biased. To overcome this issue, we demonstrate how to define a suitable ``physical distance" between system's units in terms of the flow they exchange \rev{across least resistance} pathways. We also show that the quantification of system efficiency might vary dramatically if flows are not adequately accounted for.


\section*{Results}

\heading{Flow exchange in complex topologies} Let us consider a complex network $G=(V, E)$, whose weighted adjacency matrix $\mathbf{W} = \{w_{ij}\}_{i, j\in V}$ characterizes both its topology -- $w_{ij}=0$ if $i, j$ are not adjacent-- and flows.

The architecture of a complex network influences the information exchange among its units and is responsible for a rich repertoire of interaction patterns. For instance, neurons exchange electro-chemical signals and their communication dynamics is relevant for the functional organization of the brain. Similarly, human flows through different geographic areas shape the functional organization of a city, and its neighborhood, or email among individuals in an organization generate a flow of messages determining how information reaches different teams.
The trade-off between communication efficiency and its cost characterizes complex systems and their robustness to perturbation in communication dynamics \cite{hens2019spatiotemporal,harush2017dynamic}.

Even more importantly, many empirical systems are characterized by connections with heterogeneous intensities and different correlations among weighted and purely topological network descriptors are ubiquitous~\cite{barrat2004architecture}, from the human brain~\cite{bullmore2009complex,bassett2017network,avena2018communication,van2019cross} to transportation networks~\cite{latora2002boston}.
Therefore, it is essential to account for these underlying weighted architectures to gain real insights about the hidden construction principles and mechanisms used to transform, process and exchange information~\cite{latora2003economic}.

The efficiency $\epsilon_{ij}$ in the communication between two nodes $i \neq j \in V$ is \rev{assumed to be inversely proportional to their} distance $d_{ij}$~\cite{latora2001efficient}. \rev{It follows that if} $i$ and $j$ belong to different connected components, i.e., $d_{ij} = \infty$, $\epsilon_{ij} = 0$.
The global communication efficiency of the network $G$ is the average over pairwise efficiencies
\begin{equation}\label{eq:eff-2001}
E(G) = \frac{1}{N(N-1)} \sum\limits_{i \neq j \in V} d_{ij}^{-1}.
\end{equation}
\rev{The natural metric on unweighted networks is the shortest-path distance. In this case $d_{ij}^{-1} \leq 1$, implying} $0\leq E(G)\leq 1$, with equality holding when $G$ is a clique and, since each pairwise communication is direct, information propagates the most efficiently.
\rev{In case of weighted networks, distances should also account for weights and for what they stand for \cite{opsahl2010}. Furthermore, weighted distances are real valued so that, in general, $E(G)\in [0, \infty[$ and depends on the scale of the weights. For this reason, a global indicator of efficiency should be re-scaled in $[0, 1]$ considering an idealized proxy of $G$, $G_{\text{ideal}}$, having maximum efficiency.}
\rev{In \cite{latora2001efficient,latora2003economic} the authors propose} to build
$G_{\text{ideal}}$ based on pairwise physical distances $\ell_{ij}$, which are supposed (i)``to be known even if in the graph there is no edge between $i$ and $j$'', i.e. $\ell_{ij} > 0$, (ii) should fulfill the constraint $\ell_{ij} \leq d_{ij}$ for all $i, j \in V$ \rev{and (iii) should be considered along with topological information in the computation of weighted shortest-path distances $d_{ij}$}. Then, $E(G_{\text{ideal}})=\frac{1}{N(N-1)} \sum\limits_{i \neq j \in V} \ell_{ij}^{-1} \geq E(G)$ and $\frac{E(G)}{E(G_{\text{ideal}})}$ -- which is henceforth denoted by $\text{GCE}(G)$ -- is correctly normalized.
For some spatial networks -- e.g. transportation systems like the railway or infrastructures such as the power grid -- the physical distances are well-defined by the underlying geometry, for others -- among which power stations and water resources -- it might be difficult to calculate physical distances because of the lack of direct information about spatial coordinates of units.
For non-spatial systems -- such as social and socio-technical systems -- $\{\ell_{ij}\}_{ij}$ can be found as \emph{ad hoc} transformations of link strengths (weights) into connection costs. For instance, \rev{$\ell_{ij}$ could be the inverse velocity of chemical reactions along a direct connection in a biological network \cite{latora2001efficient}}, or the minimum between 1 and the inverse number of edges between $i$ and $j$ in network with multiple unweighted edges \cite{latora2003economic}.
\rev{Unfortunately, this apparently straightforward procedure hides several issues, e.g. if there is no direct connection between two bio-chemical units in a connected network, their physical distance is infinite, according to the previous definition, while their weighted shortest-path distance will be some positive real number, violating (ii).
Furthermore,} in case of real positive weight $w_{ij} \in \mathbb{R}_+$ one cannot take $\ell_{ij}= \min\{1, \frac{1}{w_{ij}}\}$, since this introduces a cut-off on weights smaller than 1.
\rev{Another widely used option~\cite{rubinov2010complex,achard2007efficiency,bullmore2011brain,watson2019braingraph,bellingeri2019}, consists in re-scaling the weights into $[0, 1]$, transforming them to costs, applying Dijkstra's algorithm for distances and then computing the efficiency by \eqref{eq:eff-2001}, without any further comparison with a $G_{\text{ideal}}$. For instance, let us mention the max-normalization of weights $\tilde{w}_{ij}=\frac{w_{ij}}{\max\limits_{i,j} \{w_{ij}\}}$, which leads to $E^{\text{MN}}(G)= \frac{E(G)}{\max\{w_{ij}\}}$ and will be used for comparison in the rest of this study.}
It follows that, in a broad spectrum of scenarios of practical interest for applications, there is no general recipe to compute $E(G_{\text{ideal}})$. \rev{See the Supplementary Materials (SM) for details.}

\begin{figure}[!t]
\centering
\includegraphics[width=.75\textwidth]{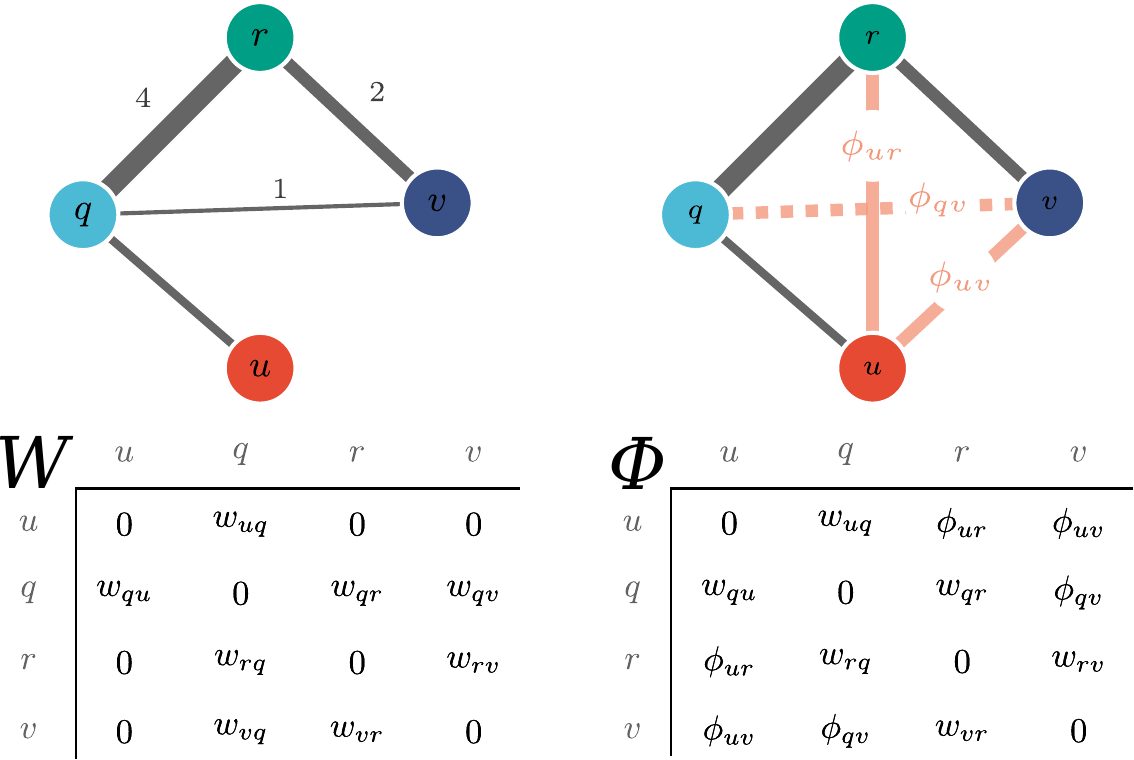}
\caption{\label{fig:flows} Computing physically-grounded ideal flows. Left: a simple weighted graph $G$ and its weighted adjacency matrix $\mathbf{W}$. Right: the artificial flows added to $G$ and their matrix $\boldsymbol{\Phi}$. $G_{\text{ideal}}$ is characterized by $\mathbf{W}_{\text{ideal}}=\frac{\mathbf{W} + \boldsymbol{\Phi}}{2}$. If the shortest path between two nodes coincides with the edge connecting them, as for $u$ and $q$, then their edge weight in $G_{\text{ideal}}$ is the same as in $G$, $\phi_{uq} = w_{uq}$. Non adjacent vertices in $G$ (e.g. $u, r$) are connected in $G_{\text{ideal}}$ by an edge with weight proportional to the elements of $\boldsymbol{\Phi}$ ($w^{\text{ideal}}_{ur}=\frac{\phi_{ur}}{2} = \frac{w_{uq} + w_{qr}}{2}$). Finally, there may be pairs of nodes with very weak connections, where longer paths have smaller weighted distances, as for $q$ and $v$ $\frac{1}{4} + \frac{1}{2} < 1$. In this case $d_{qv}=\frac{3}{4}$ and $\phi_{qv} = \rev{6}$ (dashed edge).}
\end{figure}

\heading{Rethinking efficiency of information flow in weighted architectures} To overcome the above issues, we build $G_{\text{ideal}}$ from the weighted graph $G$ in such a way that physical distances are not necessarily calculated from metadata or accessible spatial information (see Fig.\ref{fig:flows}).

We assume hereafter that edge weights are non-negative and represent the \rev{strength} of connections.
Recall that a \emph{path} is the sequence of vertices in a non-intersecting walk across the network; the \emph{length of the path} is the number of edges in -- or the sum of weights along -- that path.
Weighted shortest-path distances are then computed minimizing the sum of the reciprocals of weights~\cite{newman2001scientific,brandes2001faster}, which can be seen as costs, over all paths between node pairs\rev{\footnote{Observe that other weighted metrics may be used, see \cite{opsahl2010}}}.
Let us denote by $SP(i, j)$ a weighted and directed shortest-path from $i$ to $j$; \rev{its length}, $d_{ij} = \sum\limits_{n, m \in SP(i, j)} w_{nm}^{-1}$, is the shortest-path distance between $i$ and $j$, while $\phi_{ij} = \sum\limits_{n, m \in SP(i, j)} w_{nm}$ is the total flow along $SP(i, j)$.

\rev{The matrix $\mathbf{\Phi} = \{\phi_{ij}\}_{i, j \in V}$ represents an artificial connectivity made of shortcuts, where total flows along shortest-paths are delivered in one topological step.}
$G_{\text{ideal}}$ is then obtained averaging between the true structure $\mathbf{W}$ and the artificial connectivity, i.e., $\mathbf{W}_{\text{ideal}} = \frac{\mathbf{\Phi} + \mathbf{W}}{2}$\footnote{Another (stronger) option would be $\mathbf{W}_{\text{ideal}} = \mathbf{\Phi}$, see SM.}.
We finally define $\ell_{ij} = \left\{(w^{\text{ideal}}_{ij})^{-1}\right\}$.

\begin{figure*}[!htp]
\centering
\includegraphics[width=\textwidth]{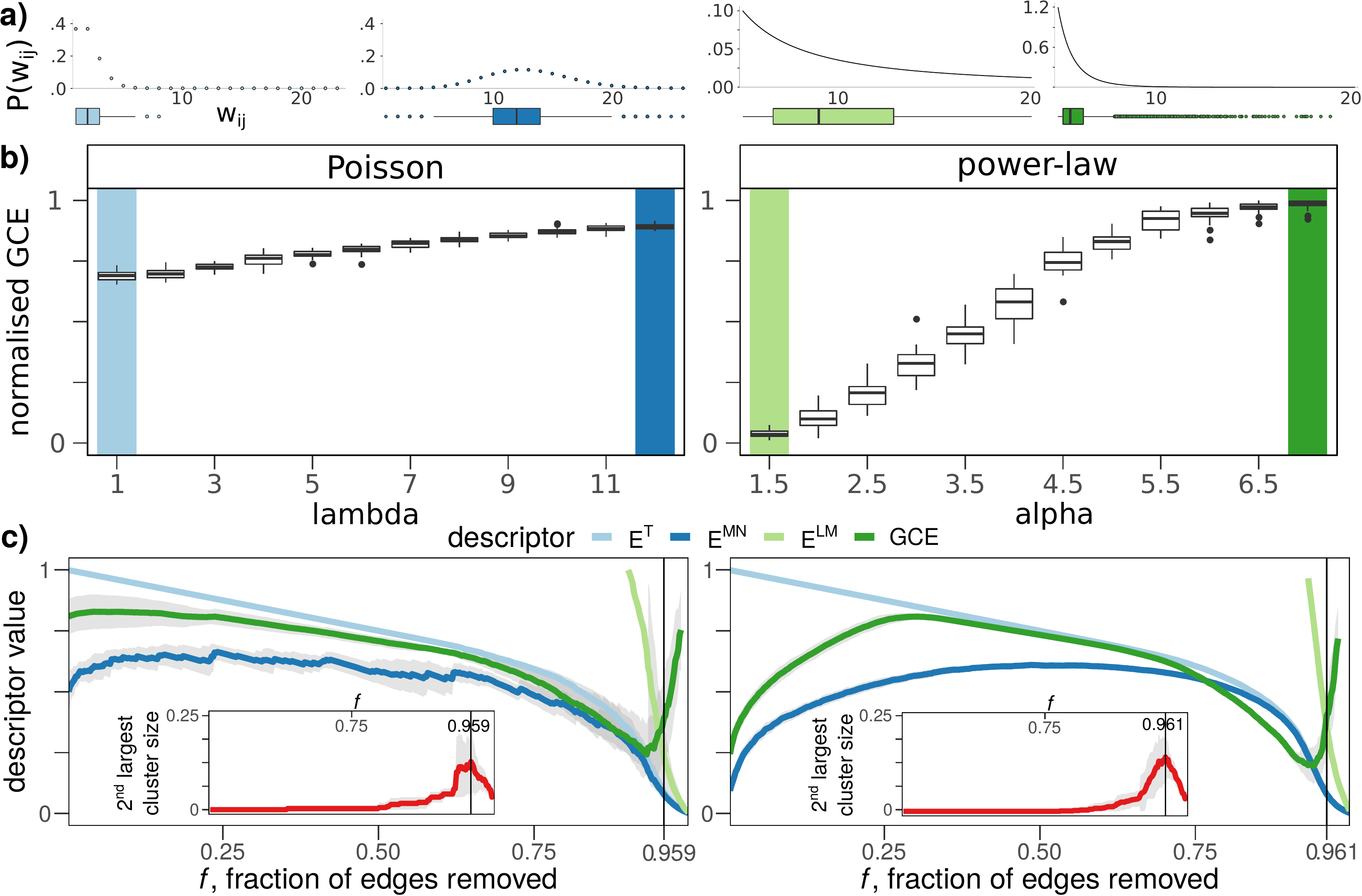}
\caption{\label{fig:synthetic}Communication efficiency of a full network with homogeneous (left) and heterogeneous (right) flows. \textbf{a)} Probability mass/density functions of edge weights for distributions and parameters highlighted in b). \textbf{b)} Global communication efficiency as a function of the free parameter of the Poisson (left) and power-law (right) distributions. As the tails of the distributions become less heavy, i.e. heterogeneity decreases, the GCE tends to $1 = E^{\text{T}}(\text{clique})$. \textbf{c)} \rev{Targeted bond percolation of an ensemble of synthetic networks  with $\mathcal{P}(2)$ (left) and power-law$(2.5)$ (right) edge weights. The shaded areas indicate standard deviations from the mean descriptor value. The average efficiency benefits from the removal of heavy links which forces the network to re-arrange its paths. The insets show the behavior of the size of the second largest connected component for $f \in [0.5, 1]$. Values on the $y-$axis have been cut to the range $[0, 1]$ -- hence only the tail of $E^{\text{LM}}$ is here visible (full plot in the SM).}}
\end{figure*}


When $G$ is connected, $G_{\text{ideal}}$ is completely connected and $\ell_{ij}$ is finite $\forall i \neq j$.
If otherwise $G$ is not connected, $G_{\text{ideal}}$ will be disconnected as well. If there is no path between $i, j$ both $\ell_{ij}=d_{ij}=\infty$ and their pairwise efficiency contribute neither to $E(G)$ nor to $E(G_{\text{ideal}})$.
Note that in this case we are computing the average communication efficiency, a global indicator, of disconnected sub-networks, which may not be meaningful.
Finally, it is possible to prove (using the Cauchy--Schwarz inequality, see SM) that the constraint $\ell_{ij} \leq d_{ij} \forall i\neq j$ is always satisfied, hence $\{\ell_{ij}\}_{ij}$ are well-defined physical distances that can be calculated for any weighted systems.
Having defined the mathematical tools, we now analyze some synthetic networks \rev{with a tunable structure}. This enables us to separate the effects of topology and flows on the global efficiency of the network.

\heading{Global efficiency of synthetic networks}
We generate two ensembles of networks with the same topology -- cliques with $30$ nodes -- and edge weights sampled from different probability distribution families.
\rev{The topological efficiency $E^{\text{T}}$, i.e., \eqref{eq:eff-2001} computed ignoring weights, is 1 for all networks, since they are cliques. We therefore focus on the weighted descriptors $E^{\text{LM}}, E^{\text{MN}}$ and GCE.}
The trivial case, $w_{ij} = w > 0$ constant, leads to $\text{GCE}=1$.
We impose more realistic homogeneous flows sampling from a Poisson distribution $\mathcal{P}(\lambda)$ with varying $\lambda$ -- Fig.\ref{fig:synthetic}(left). Since zero belongs to the support of the distribution, we add one to each sample to keep the complete connectedness of the network.
The heterogeneity in the weighted structure is instead modeled with $w_{ij}$ following power-laws$(\alpha)$ with a lower bound $x_{min}=5$~\cite{clauset2009power} -- Fig.\ref{fig:synthetic}(right).
In panel a) we show some probability mass (resp. density) functions as the free parameters $\lambda$ and $\alpha$ increase: the variance increases for the Poisson and decreases for the power-law.
\rev{Another index of heterogeneity is the kurtosis (tailness) of the distribution, which decreases for both.}
Panel b) shows the global communication efficiency, evaluated on 30 random samples for each distribution, as a function of $\lambda$ (resp. $\alpha$).
\rev{These networks are topologically equally efficient,} however, accounting for the weights can lead to dramatically different results.
The extreme heterogeneity of edge weights, characteristics of power-law distributions with small scaling exponent, strongly reduces the average communication efficiency of the network.
Furthermore, as the tails of the weight distributions become lighter the weighted GCE tends to the topological one.

We next study the interplay between weights heterogeneity and topology through bond percolation~\cite{latora2005vulnerability,bellingeri2019}.
By removing edges in decreasing weight order, we trim the tail of the weights distributions, reducing their heterogeneity.
\rev{In Fig.\ref{fig:synthetic}~c) we plot the four efficiency quantifiers as functions of the fraction $f$ of removed edges and averaged over 30 random realizations of each model. Shaded areas indicate the standard deviation from the mean.
We denote by $G_f$ the damaged network obtained from $G$ removing $f\%$ of its heavier links.
$G_0$ is, topologically, a clique, so $E^{\text{T}}(G_0)=1$. In $E^{\text{T}}(G_f)$ the denominator is always $N(N-1)$, hence $G_f$ is compared with a clique, by definition and $E^{\text{T}}(G_f)$ decreases monotonically. On the other side $E^{\text{MN}}$, $E^{\text{LM}}$ (SM, Fig.S3) and GCE use the flows of $G_f$ to build the corresponding $G_f^{\text{ideal}}$ and are, consequently non-monotone functions of $f$. It might seem a limitation\footnote{to overcome it in percolation applications, we propose a modification of the GCE in the SM} nevertheless,} it allows us to compare a series of networks $G_f$ with slightly different topologies and \rev{flows that become increasingly homogeneous.}
\rev{$E^{\text{MN}}$ and GCE behave similarly, although $E^{\text{MN}}$ has larger fluctuations because at each step the edge with maximum weight is removed.}
As expected, there are clear differences in the percolation plots of Poissonian and power-law network flows, but in both cases removing the heaviest links produces an increase in the average communication efficiency.
In both cases, when the flows become more homogeneous the GCE depends largely on the topology,
Finally, when the network is disrupted -- near the critical threshold $f_c$ indicated by the maximum of the second largest cluster size (insets of Fig.\ref{fig:synthetic}) -- the GCE has a break-down point\rev{, since we are averaging the efficiencies of many, small, distinct (and maybe efficient) disconnected networks. More details in the SM.}

\begin{figure}[!b]
\centering
\includegraphics[width=0.75\textwidth]{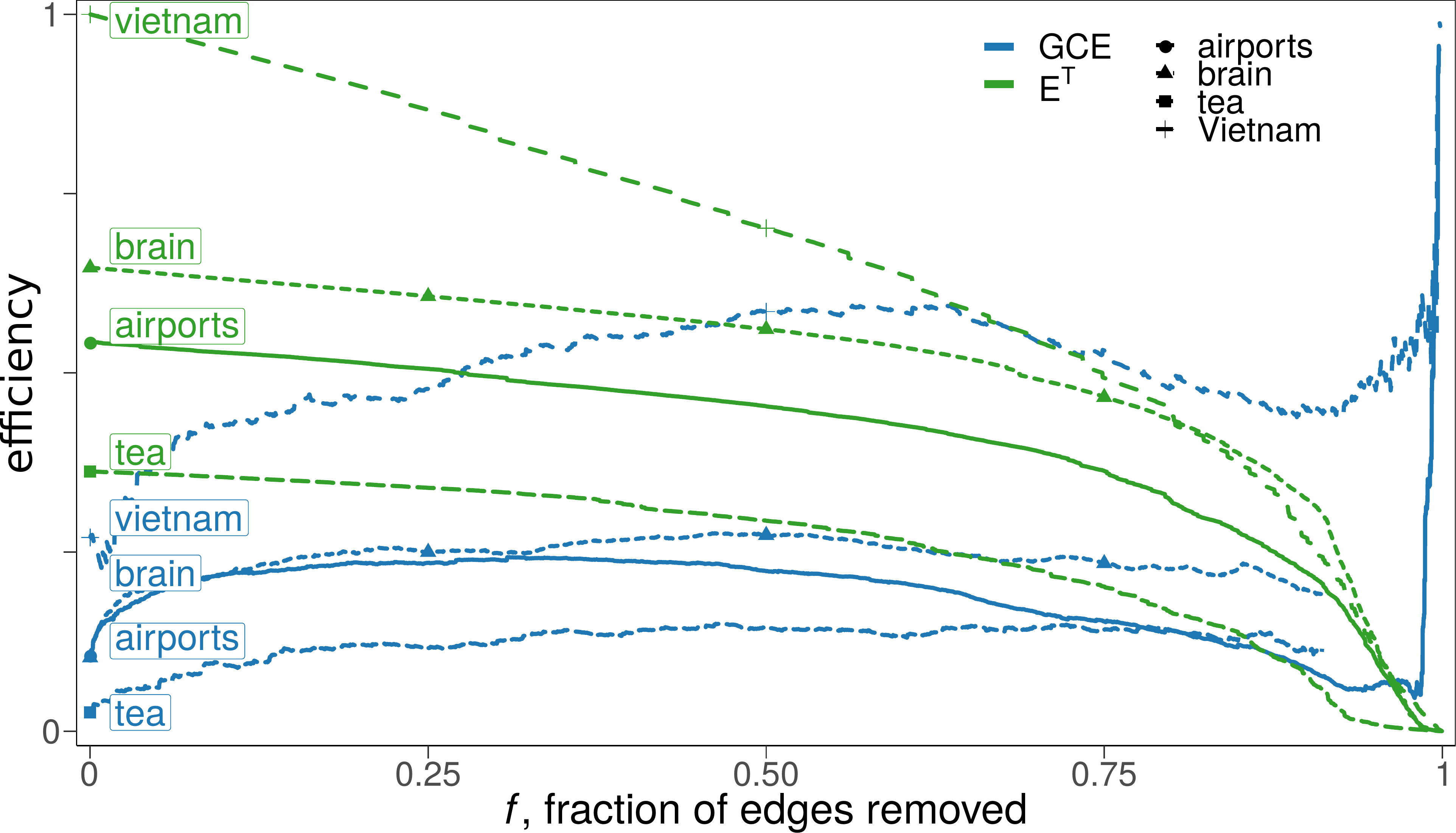}
\caption{\label{fig:comparison}Targeted bond percolation of real interconnected systems - $E^{\text{T}}(G_f)$ and GCE$(G_f)$. See SM Fig.S5-S7.}
\end{figure}

\heading{Global efficiency of real interconnected systems}
We use our framework to study the efficiency of four real systems (see Tab.~\ref{tab:datasets}).  Figure~\ref{fig:comparison} shows the curves corresponding to \rev{$E^{\text{T}}(G_f)$ and $\text{GCE}(G_f)$}.

From the FAO worldwide food trade network we selected the layers of cocoa, coffee, tea, and tobacco. From the migration dataset we selected internal migration flows inside three Asian regions: India, China and Vietnam. From the worldwide air traffic network we extracted the traffic in and between Europe and Africa. Finally, we consider the structural connectivity of human brain -- quantified through diffusion tensor imaging (DTI) and fiber tractography methods.

\begin{table}[!hb]
\centering
\begin{tabular}{lllll}
Dataset        &             & $|V|$ & $|E|$ & Ref.    \\\hline
FAO            & cocoa       & 159 & 2,081  & \cite{de2015structural}\\
               & coffee      & 184 & 7,760  & \\
               & tea         & 172 & 3,297  & \\
               & tobacco     & 183 & 3,623  & \\\hline
Migrations     & China       & 30 & 870    & \cite{worldpop,Sorichetta2016}\\
               & India       & 32 & 992    & \\
               & Vietnam     & 63 & 3,906   & \\\hline
Transportation & airports    & 299 & 12,919 & \cite{brockmann2013hidden}\\\hline
Biological     & human brain & 188 & 10,836 & \cite{brown2012ucla}
\end{tabular}
\caption{Real flow networks and corresponding scales. Multiple edges have been aggregated and loops removed.}
\label{tab:datasets}
\end{table}

These real networks have different properties, among which edge density and weight distribution.
Independently from the system, ignoring the network flows leads to an overestimation of the average efficiency, especially when flows are highly heterogeneous.
The network of internal migration, is the most efficient, but it also has the highest cost being a clique.
The tea trade network is the most inefficient. Finally the brain and the airports network have similar GCEs until the first 25\% of their edges are removed, with the brain remaining afterwards more efficient w.r.t the reduced flows.
Observe that the total flow could be restored, while keeping a specific efficiency value, re-distributing the removed flow on the remaining links.
In general, removing those edges monopolizing shortest-paths forces \rev{their reallocation inducing} an increase of the global weighted efficiency.

\section*{Discussion}

Exchanging information is one of the main functions and \emph{raison d'\^etre} of many real complex systems and quantifying how efficiently they perform this task is of great interest for different disciplines. Consequently, the concept of communication and transport efficiency is relevant for a broad range of applications, from public transportation to the human brain and the Internet. Such networks are indeed flow networks: links are not only weighted but they encode volumes (of people, electro-chemical junctions, packets, so forth and so on). While there is a widely adopted descriptor for the global communication efficiency in case of unweighted networks, we have found that its generalization to the case of weighted network might not be suitable in general. In this work we have identified and well understood the current mathematical limitations of the current measure.

A direct consequence of our analysis is that an estimation of global efficiency can be trusted only under specific conditions: i.e., the analysis of efficiency in the case of real network flows can not be performed or, alternatively, when it is performed it might lead to important underestimation or overestimation of results.  Since flow networks are ubiquitous, here we have proposed the most general definition of the global communication efficiency for weighted directed networks, which does not assume any other (meta-)information on the system.

Using our physically grounded definition of flow network efficiency, our results indicate that one can achieve a desired level of efficiency by wisely redistributing weights, instead of altering the underlying topology. This result is relevant for practical applications, since it is not always guaranteed that one can rewire or dramatically change with other interventions the network connectivity. In fact, altering network structure is usually expensive in economic or energetic terms. Conveniently, our framework works under mild assumptions about the underlying topology \rev{and about the ideal and most efficient network}, with no metadata, nor additional spatial (e.g., geographic) information on the system, allowing for trustworthy applications to empirical problems. Remarkably, the framework allows for a complementary view of bond percolation from a functional perspective, allowing us to gain new insights about critical phases of information exchange and network flows in additions to topological ones.


\section*{Methods}

\heading{Mathematical details on the normalizing procedure}
We provide the proof of $d_{ij} \geq \ell_{ij} ~\forall i\neq j \in V_G$, which is sufficient for the GCE to be correctly normalized in $[0, 1]$.
Recall that $SP(i, j)$ denotes a weighted (directed) shortest-path from $i$ to $j$ and $d_{ij} = \sum\limits_{n, m \in SP(i, j)} w_{nm}^{-1}$.
Observe also that, if the shortest-path between $i, j$ coincides with their link $(i, j)$ the number of vertices in the sequence is $|SP(i, j)| = 2$ and their shortest-path distance is $d(i, j) = \frac{1}{w_{ij}}$.
The total flow between $i$ and $j$ through the shortest-path $SP(i, j)$ is defined as $\phi_{ij} = \sum\limits_{n, m \in SP(i, j)} w_{nm}$.

Before proving our main statements we write an inequality, which will be extensively used in the following proofs.
The Cauchy--Schwarz inequality for vectors $\mathbf{u}, \mathbf{v}$ in an inner product space reads
$|\langle\mathbf{u}, \mathbf{v}\rangle|^{2}~\leq~\langle\mathbf{u}, \mathbf{u}\rangle \cdot\langle\mathbf{v}, \mathbf{v}\rangle$.
Taking $\mathbf{u} = \left(\frac{1}{\sqrt{x_1}}, \dots, \frac{1}{\sqrt{x_n}}\right)$ and $\mathbf{v} = \left(\sqrt{x_1}, \cdots, \sqrt{x_n}\right)$ the inequality becomes
\begin{align}
n^2 = \left( \sum\limits_{i=1}^n\frac{\sqrt{x_i}}{\sqrt{x_i}} \right)^2 & \leq  \left(\sum\limits_{i=1}^n \frac{1}{x_i}\right) \left( \sum\limits_{i=1}^n x_i \right)\nonumber\\
n^2 \left( \sum\limits_{i=1}^n x_i \right)^{-1} & \leq \left(\sum\limits_{i=1}^n \frac{1}{x_i}\right). \label{eq:CS-ineq}
\end{align}
\eqref{eq:CS-ineq} states that for non-negative real numbers $x_1, \dots, x_n$ the inverse of their sum is smaller or equal to the sum of their reciprocals.

Since we have assumed edges weights to be positive we can apply the inequality, which leads us to

\begin{align}
\left(\sum\limits_{n, m \in SP(i, j)} w_{nm} \right)^{-1} & \leq |SP(i, j)|^2 \left(\sum\limits_{n, m \in SP(i, j)} w_{nm} \right)^{-1} \leq \sum\limits_{n, m \in SP(i, j)} w_{nm}^{-1}.\label{eq:CS-weights}
\end{align}
Observe that $|SP(i, j)|\geq 2$ if $G$ is connected, therefore the first inequality is actually strict.

From \eqref{eq:CS-weights} we can derive useful inequalities involving $w_{ij}, \phi_{ij}, d_{ij}$ and $\ell_{ij}$:
\begin{align}\label{eq:phi-dij}
    \phi_{ij}^{-1} & = \left(\sum\limits_{n, m \in SP(i, j)} w_{nm} \right)^{-1} \leq \sum\limits_{n, m \in SP(i, j)} w_{nm}^{-1} = d_{ij}
\end{align}
note that if $w_{ij}\neq 0$, it also holds $d_{ij} \leq \frac{1}{w_{ij}}$.

It is also possible to prove that $\phi_{ij} \geq w_{ij},~\forall i, j \in V_G$. Indeed, if $i, j$ are not adjacent then $w_{ij}=0$ but, since $G$ is connected, there is a path between them with $\phi_{ij}>0$.
If instead, they are adjacent, either $\phi_{ij} = w_{ij}$ meaning that the weighted shortest-path coincides with the edge $(i, j)$, or there is a shortest-path going through other vertices, such that $d_{ij} = \sum\limits_{n, m \in SP(i, j)} w_{nm}^{-1}<\frac{1}{w_{ij}}$ and the claim follows from \eqref{eq:phi-dij}.

Starting from the definition of physical distances $\ell_{ij}$, using simple inequalities and \eqref{eq:CS-weights}
\begin{align}
    \ell_{ij} & =  2 \left(w_{ij} + \phi_{ij}\right)^{-1} \leq 2 \left(\sum\limits_{n, m \in SP(i, j)} w_{nm} \right)^{-1} \leq |SP(i, j)|^2 \left(\sum\limits_{n, m \in SP(i, j)} w_{nm} \right)^{-1} \nonumber\\
    & \leq \sum\limits_{n, m \in SP(i, j)} w_{nm}^{-1} = d_{ij}.\label{eq:ineq-ld}
\end{align}
Again, for a connected network $G$ the strict inequality $\ell_{ij} < d_{ij}$ holds.

Finally, $\phi_{ij} = 0$ if and only if $i, j$ lie in disconnected components and the ideal network will be disconnected as the original one.
In this case both $d_{ij}=\frac{1}{\phi_{ij}}=\infty$ and the missing links among disconnected components will not produce an under-estimation of the efficiencies of the subgraphs.
Of course, if the network is very fragmented the GCE, a global descriptor, will not be very informative. Below, we propose a variant of the GCE, which is most appropriate in this case and in percolation simulations in general, see Suppl. Material.

\heading{Comparisons with other weighted efficiency measures.}
In this work we introduced existing measures of topological and weighted efficiency, more specifically, $E^{\text{T}}$ the topological efficiency defined in \cite{latora2001efficient}, $E^{\text{LM}}$ defined by Latora and Marchiori in \cite{latora2003economic}, $E^{\text{MN}}$ obtained evaluating the efficiency on the network with max-normalized weights.

Let us recall the definition of efficiency \cite{latora2001efficient}
\begin{equation}\label{eq:eff-2001}
E(G) = \frac{1}{N(N-1)} \sum\limits_{i \neq j \in V} d_{ij}^{-1}.
\end{equation}
where we have the parallel sum of pairwise distances $\sum\limits_{i \neq j \in V} d_{ij}^{-1}$ divided by the number of non-diagonal entries in the distances matrix, i.e., $N(N-1)$. In the topological case this last term plays the role of a normalizing factor, since the parallel sum of shortest-path distances in a clique is exactly equal to $N(N-1)$.
We refer to the efficiency \eqref{eq:eff-2001} evaluated without edge weights, or in other words with topological shortest-path distances $\{d_{ij}\}_{i,j}$, as \textit{the topological efficiency} and indicate it by $E^{\text{T}}$. Then $E^{\text{T}}$ naturally lies in $[0, 1]$.

The main difficulty arising in the definition of a weighted efficiency descriptor, have to do with the diversity of information that can be encoded as edge weights in a network.
Usually weights represent connection strengths and connection costs are obtained as a function (e.g., inverse) of weights.
Given the connection costs one can compute weighted shortest-path distances \cite{dijkstra1959note,newman2001scientific,brandes2001faster,opsahl2010}, which vary in $[0, +\infty]$ and therefore $E(G) \in [0, +\infty)$ and needs to be re-scaled (or normalized) in order to be comparable among different systems.

$E^{\text{MN}}$, which has been used, for instance, in \cite{rubinov2010complex,bullmore2011brain,watson2019braingraph,bellingeri2019} is the simplest generalization of $E^{\text{T}}$ to the weighted case: re-scaling the weights to $[0, 1]$ implies that shortest-path distances $d_{ij} \geq 1$, since weighted shortest-paths are those paths minimizing the sum of edge costs, that is, inverse weights. Consequently, being $d_{ij} \geq 1$, \eqref{eq:eff-2001} results to be normalized.
Different re-scaling transformations of weights are possible, the most common is the max-normalization (from which the superscript MN) $\tilde{w}_{ij}=\frac{w_{ij}}{\max\limits_{i,j} w_{ij}}$.
The cost of edges is then $\tilde{w}_{ij}^{-1}$.
We show, now, that $E^{\text{MN}}(G) = \frac{E(G)}{w_{max}}$, where $E(G)$ is \eqref{eq:eff-2001} calculated on weighted geodesic distances without the max-normalization of weights.
Let $w_{\text{max}}$ be the maximum weight over all edges of a weighted network $G=(V, E)$ and let $SP(i, j)$ be a weighted shortest path between $i, j \in V$.
Observe that the max-normalization of weights does not affect the shortest path, but it does affect the sp-distance
\begin{align*}
    \tilde{d}_{ij} & = \sum_{n, m \in SP(i, j)} \frac{1}{\tilde{w}_{ij}}  = \sum_{n, m \in SP(i, j)} \frac{w_{\text{max}}}{w_{ij}}  = w_{\text{max}} d_{ij}.
\end{align*}
Finally,
\begin{align*}
    E^{\text{MN}}(G) & = \frac{1}{N(N-1)} \sum\limits_{i \neq j \in V} \tilde{d}_{ij}^{-1}  \frac{1}{w_{\text{max}}} E(G).
\end{align*}

Of course, not only the max-normalization and inverse are available, for instance, in \cite{achard2007efficiency} weights are wavelet correlation coefficient between regions in the brain and the cost of the connection between regions $i$ and $j$ is defined as $c_{ij} = 1 - w_{ij}$.

$E^{\text{LM}}$ is the weighted generalization of $E(G)$ proposed by Latora and Marchiori in \cite{latora2003economic}. The idea is to normalize $E(G)$ considering an ideal case $G_{\text{ideal}}$, where all possible edges are present in the idealized graph and the information propagated most efficiently. Then,
\begin{equation}\label{eq:ideal-case}
    \frac{E(G)}{E(G_{\text{ideal}})} = \frac{ \frac{1}{N(N-1)} \sum\limits_{i \neq j \in V} d_{ij}^{-1}}{ \frac{1}{N(N-1)} \sum\limits_{i \neq j \in V} l_{ij}^{-1}} \leq 1.
\end{equation}
Observing that a sufficient condition for \eqref{eq:ideal-case} is $0 \leq \ell_{ij} \leq d_{ij}$ for all $i, j, \in V$, defining $G_{\text{ideal}}$ reduces to building the matrix $\{\ell_{ij}\}_{i,j}$.
They called $\ell_{ij}$ physical distances, in contrast to shortest-path distances, highlighting that the latter are computed using "the information contained both in the binary adjacency matrix and in $\{\ell_{ij}\}_{i,j}$". Observe that the matrix the matrix $\{\ell_{ij}\}_{i,j}$ is, in every respect, a matrix of connection costs.
In \cite{latora2003economic} (Sec.~3) the authors give some examples to $\ell_{ij}$ from edge weights. For instance, if weights $w_{ij}\geq 1$ one can define $\ell_{ij} = \min\left\{1, \frac{1}{w_{ij}}\right\}$, which is the transformation adopted in this work to compute $E^{\text{LM}}$.

We refer to the Suppl. Material (S1 and S2) for the full plot corresponding to Fig.2 b) of this study. The GCE converges faster to 1 as the weight distributions become less heterogeneous (in terms of kurtosis). We claim that the maximum of the GCE is obtained not only for full networks with constant edge weight distribution, but it is sufficient to have a uniform edge betweenness, as shown in S2.

Finally, the panel c) of Fig.~2, without the cut on the range of $y-$values is reported in Fig.~S3.
The percolation simulation consists in removing edges from an undirected weighted full network $G$, in decreasing weight order.
We indicate by $f$ the fraction of edges removed from $G$ and by $G_f$ the resulting, damaged, network with $G_0 = G$. We then evaluate the efficiency of $G_f$ by means of the already described measures: $E^{\text{T}}, E^{\text{MN}}, E^{\text{LM}}$, GCE.
We repeat the process 30 times, sampling the edge weights from a Poisson distribution with parameter $\lambda = 2$, Fig.~S3(top), and 30 times,  sampling the edge weights from a power-law distribution with free parameter $\alpha = 2.5$, Fig.~S2(bottom).
We include also the plots for common percolation indicators, such as the total weight of the largest connected component (LCC) re-scaled in $[0, 1]$, the size of the second largest LCC -- divided by $N = |V|$ -- and the number of clusters -- also divided by $N = |V|$; see Fig.~S4.

Our normalization procedure can also be used to build a slightly modified version of the GCE that plays the role of a weighted integrity descriptor for percolation analysis.
Let $G_0^{\text{ideal}}$ be the idealized network corresponding to $G_0$ build as described in our study.
Then
\begin{equation}
    \text{GCE}^*(G_f) = \frac{E(G_f)}{E(G_0^{\text{ideal}})}
\end{equation}
is normalized in $[0, 1]$ and it is a monotone decreasing function w.r.t. $f$.

This variant of the GCE will be evaluated for real networks in the last section of the SM.

\heading{On artificial flows} We choose to build the artificial flows integrating weights over paths, but this is not the only possibility, provided that constraint $\ell_{ij}\leq d_{ij}$ is satisfied.
Now, using the same definitions and notation adopted in this study,
\begin{align}
d_{ij} & = \sum_{n,m \in SP(i, j)} \frac{1}{w_{nm}}\nonumber\\
& \geq \frac{1}{\min\{w_{nm}: n, m \in SP(i, j)\}} \label{eq:gce-min}\\
& \geq \frac{1}{\max\{w_{nm}: n, m \in SP(i, j)\}} \label{eq:gce-max}\\
& \geq \frac{1}{\phi_{ij}} \nonumber\\
& \geq \frac{1}{\max\{w_{nm}: n, m \in  V\}}. \nonumber
\end{align}
So both the minimum (or maximum) weight over the path are valid choices, as well as, the sum (our choice) and the maximum over all edges (the already discussed max-normalization).
Now, when using the sum fo flows over paths, we can combine the two sources of information $\mathbf{W}$ and $\mathbf{\Phi}$ through the arithmetic mean, while this strategy is not possible if we define $\phi^*_{ij} = \min\{w_{nm}: n, m \in SP(i, j)\}$ (resp. $\max$) because we cannot prove that $\ell_{ij} = \frac{2}{\phi^*_{ij} + w_{ij}} \leq d_{ij}$, so we should drop $\mathbf{W}$ and simply define $\mathbf{W}_{\text{ideal}} = \mathbf{\Phi^*}$.
We implemented these other choices and show the results on our synthetic networks ensembles in Fig.~S5.
The two variants, called here $\text{GCE}^{\text{min}}$, $\text{GCE}^{\text{max}}$, converge faster to 1, since both minimizations \eqref{eq:gce-min}-\eqref{eq:gce-max} are less strict than ours (the sum). Taking the minimum \eqref{eq:gce-min}, in particular, may result in values of efficiency spanning a narrow range, near 1, with a consequent difficulty in distinguishing networks on the basis of efficiency. Furthermore, in the bottom panel of Fig.~S5 we can see a decreasing-increasing behavior of $\text{GCE}^{\text{min}}$ which is, from our point of view, not desirable.
$\text{GCE}^{\text{max}}$ displays, in general, a larger variability.
We opt for the sum, since it has a physical meaning in terms of total flow of a sub-graph (a path $SP(i, j)$), it allows us to average the artificial flows matrix with the original flows given by $\mathbf{W}$ and, last but not least, it is easily worked with in mathematical terms (simplifies rigorous proofs).

\heading{On the normalized weighted efficiency of Latora and Marchiori \cite{latora2003economic}}
Let us take $G$ as the subgraph consisting of vertices $q, r, v$ of Fig.~1 and suppose that the weights are the result of the aggregation of multiple binary connections. Its weighted adjacency matrix is
\begin{equation*}
    \mathbf{W} = \begin{pmatrix}
    \cdot & 4 & 1 \\
    4 & \cdot & 2 \\
    1 & 2 & \cdot
    \end{pmatrix}
\end{equation*}
We can compute physical distances $\ell_{ij}$ following the suggestions in \cite{latora2003economic}
\begin{equation*}
    \mathbf{L} = \begin{pmatrix}
    \cdot & \frac14 & 1 \\
    \frac14 & \cdot & \frac12 \\
    1 & \frac12 & \cdot
    \end{pmatrix}
\end{equation*}
and shortest-path distances minimizing the sum of costs (i.e. inverse weights)
\begin{equation*}
    \mathbf{D} = \begin{pmatrix}
    \cdot & \frac14 & \frac34 \\
    \frac14 & \cdot & \frac12 \\
    \frac34 & \frac12 & \cdot
    \end{pmatrix}
\end{equation*}
The global communication efficiency defined in \cite{latora2003economic} is given by $E_{\mathrm{glob}}=\frac{E(G)}{E\left(G_{\mathrm{ideal}}\right)}$, where $E(G)=\frac{1}{N(N-1)} \sum\limits_{i \neq j} \frac{1}{d_{ij}}$ and $E(G_{\mathrm{ideal}})=\frac{1}{N(N-1)} \sum\limits_{i \neq j} \frac{1}{\ell_{ij}}$.
Observe that the condition (which is sufficient for $\frac{E(G)}{E(G_{\mathrm{ideal}})} \leq 1$)
\begin{equation}\label{eq:dl-condition}
    d_{ij} \geq \ell_{ij} \quad \forall i\neq j \in V_G
\end{equation}
is not satisfied for $i=1, j=3$ and this causes $\text{GCE}=\frac{E(G)}{E(G_{\mathrm{ideal}})}= \frac{22}{9}\left(\frac{7}{3}\right)^{-1} > 1$.

This counter-example on the statement \eqref{eq:dl-condition} is not a pathological case: \eqref{eq:dl-condition} is violated whenever the weighted shortest-path between adjacent nodes $i, j$ does not traverse the direct link $e_{ij}$, i.e. $d_{ij} < \frac{1}{w_{ij}}$ and it may often happen in real networks with large heterogeneous weights.

Trying to reproduce the results in \cite{latora2003economic}, we considered the neural network of the \emph{C. elegans} \cite{watts1998collective,latora2003economic}, with data from \url{http://www-personal.umich.edu/~mejn/netdata/}.
Firstly, we aggregate multiple edges, obtaining a simple, directed, weighted network with $N=297$ nodes, $m = 2345$ edges and weights in the range $[1, 70]$.
If we consider the network as undirected, we obtain $m = 2148$ edges and weights in the range $[1, 72]$. The data are not the same used in \cite{latora2003economic}, so we cannot reproduce their results exactly. Let us focus on the undirected network: Fig.~S6 shows the distance matrix $D$ evaluated using Dijkstra's algorithm with the reciprocal of edge weights, and the matrix of physical distances $L$, with $\ell_{ij} = \min\{1, \frac{1}{w_{ij}}\}$.

\heading{Real interconnected systems, additional results} Here we first apply the variant of the GCE, i.e., GCE$^*(G_f)$ to the networks of migrations inside Vietnam and of human brain; secondly, we report the detailed percolation results for the real network flows discussed in this study.

We refer to Suppl. Material to show the behavior of $\text{GCE}^*(G)$ -- both topological and weighted -- for two of the real networks from Tab.1.

Finally, we show the percolation plots for the remaining datasets studied in this work, see Figures~S9.

\heading{Contributions} GB performed the theoretical analysis, the numerical experiments, the data analysis and wrote the paper. RG performed the numerical experiments and wrote the manuscript. MDD conceived and designed the study and wrote the manuscript.

\heading{Competing financial interests} The authors declare no competing financial interests.

\heading{Acknowledgements} The authors thank Dirk Brockmann for providing us with the worldwide air-transportation flow data.

\bibliographystyle{naturemag}

\begin{small}

\end{small}



\end{document}


\title{Supplementary material: Quantifying efficient information exchange in real network flows}

\author{Giulia Bertagnolli}
\affiliation{CoMuNe Lab, Fondazione Bruno Kessler, Via Sommarive 18, 38123 Povo (TN), Italy}
\affiliation{University of Trento, Department of Mathematica, Via Sommarive, 14, 38123 Povo (TN), Italy}

\author{Riccardo Gallotti}
\affiliation{CoMuNe Lab, Fondazione Bruno Kessler, Via Sommarive 18, 38123 Povo (TN), Italy}

\author{Manlio De Domenico}
\email[Corresponding author:~]{mdedomenico@fbk.eu}%
\affiliation{CoMuNe Lab, Fondazione Bruno Kessler, Via Sommarive 18, 38123 Povo (TN), Italy}

\date{\today}

\maketitle

\beginsupplement

\heading{Mathematical details on the normalizing procedure}
We provide the proof of $d_{ij} \geq \ell_{ij} ~\forall i\neq j \in V_G$, which is sufficient for the GCE to be correctly normalized in $[0, 1]$.
Recall that $SP(i, j)$ denotes a weighted (directed) shortest-path from $i$ to $j$ and $d_{ij} = \sum\limits_{n, m \in SP(i, j)} w_{nm}^{-1}$.
Observe also that, if the shortest-path between $i, j$ coincides with their link $(i, j)$ the number of vertices in the sequence is $|SP(i, j)| = 2$ and their shortest-path distance is $d(i, j) = \frac{1}{w_{ij}}$.
The total flow between $i$ and $j$ through the shortest-path $SP(i, j)$ is defined as $\phi_{ij} = \sum\limits_{n, m \in SP(i, j)} w_{nm}$.

Before proving the statements in the Letter we write an inequality, which will be extensively used in the following proofs.
The Cauchy--Schwarz inequality for vectors $\mathbf{u}, \mathbf{v}$ in an inner product space reads
$|\langle\mathbf{u}, \mathbf{v}\rangle|^{2}~\leq~\langle\mathbf{u}, \mathbf{u}\rangle \cdot\langle\mathbf{v}, \mathbf{v}\rangle$.
Taking $\mathbf{u} = \left(\frac{1}{\sqrt{x_1}}, \dots, \frac{1}{\sqrt{x_n}}\right)$ and $\mathbf{v} = \left(\sqrt{x_1}, \cdots, \sqrt{x_n}\right)$ the inequality becomes
\begin{align}
n^2 = \left( \sum\limits_{i=1}^n\frac{\sqrt{x_i}}{\sqrt{x_i}} \right)^2 & \leq  \left(\sum\limits_{i=1}^n \frac{1}{x_i}\right) \left( \sum\limits_{i=1}^n x_i \right)\nonumber\\
n^2 \left( \sum\limits_{i=1}^n x_i \right)^{-1} & \leq \left(\sum\limits_{i=1}^n \frac{1}{x_i}\right). \label{eq:CS-ineq}
\end{align}
\eqref{eq:CS-ineq} states that for non-negative real numbers $x_1, \dots, x_n$ the inverse of their sum is smaller or equal to the sum of their reciprocals.

Since we have assumed edges weights to be positive we can apply the inequality, which leads us to

\begin{align}
\left(\sum\limits_{n, m \in SP(i, j)} w_{nm} \right)^{-1} & \leq |SP(i, j)|^2 \left(\sum\limits_{n, m \in SP(i, j)} w_{nm} \right)^{-1}\nonumber\\
& \leq \sum\limits_{n, m \in SP(i, j)} w_{nm}^{-1}.\label{eq:CS-weights}
\end{align}
Observe that $|SP(i, j)|\geq 2$ if $G$ is connected, therefore the first inequality is actually strict.

From \eqref{eq:CS-weights} we can derive useful inequalities involving $w_{ij}, \phi_{ij}, d_{ij}$ and $\ell_{ij}$:
\begin{align}\label{eq:phi-dij}
    \phi_{ij}^{-1} & = \left(\sum\limits_{n, m \in SP(i, j)} w_{nm} \right)^{-1} \nonumber \\
    & \leq \sum\limits_{n, m \in SP(i, j)} w_{nm}^{-1} = d_{ij}
\end{align}
note that if $w_{ij}\neq 0$, it also holds $d_{ij} \leq \frac{1}{w_{ij}}$.

It is also possible to prove that $\phi_{ij} \geq w_{ij},~\forall i, j \in V_G$. Indeed, if $i, j$ are not adjacent then $w_{ij}=0$ but, since $G$ is connected, there is a path between them with $\phi_{ij}>0$.
If instead, they are adjacent, either $\phi_{ij} = w_{ij}$ meaning that the weighted shortest-path coincides with the edge $(i, j)$, or there is a shortest-path going through other vertices, such that $d_{ij} = \sum\limits_{n, m \in SP(i, j)} w_{nm}^{-1}<\frac{1}{w_{ij}}$ and the claim follows from \eqref{eq:phi-dij}.

Starting from the definition of physical distances $\ell_{ij}$, using simple inequalities and \eqref{eq:CS-weights}
\begin{align}
    \ell_{ij} & =  2 \left(w_{ij} + \phi_{ij}\right)^{-1} \nonumber \\
    & \leq 2 \left(\sum\limits_{n, m \in SP(i, j)} w_{nm} \right)^{-1} \nonumber\\
    & \leq |SP(i, j)|^2 \left(\sum\limits_{n, m \in SP(i, j)} w_{nm} \right)^{-1} \nonumber \\
    & \leq \sum\limits_{n, m \in SP(i, j)} w_{nm}^{-1} = d_{ij}.\label{eq:ineq-ld}
\end{align}
Again, for a connected network $G$ the strict inequality $\ell_{ij} < d_{ij}$ holds.

Finally, $\phi_{ij} = 0$ if and only if $i, j$ lie in disconnected components and the ideal network will be disconnected as the original one.
In this case both $d_{ij}=\frac{1}{\phi_{ij}}=\infty$ and the missing links among disconnected components will not produce an under-estimation of the efficiencies of the subgraphs.
Of course, if the network is very fragmented the GCE, a global descriptor, will not be very informative. Below, we propose a variant of the GCE, which is most appropriate in this case and in percolation simulations in general, see Fig.\ref{fig:gce-monotone}.

\heading{Real interconnected systems, additional results} Here we report the detailed percolation results for the real network flows discussed in the Letter.
\begin{figure*}[!t]
\centering
\includegraphics[width=0.9\textwidth]{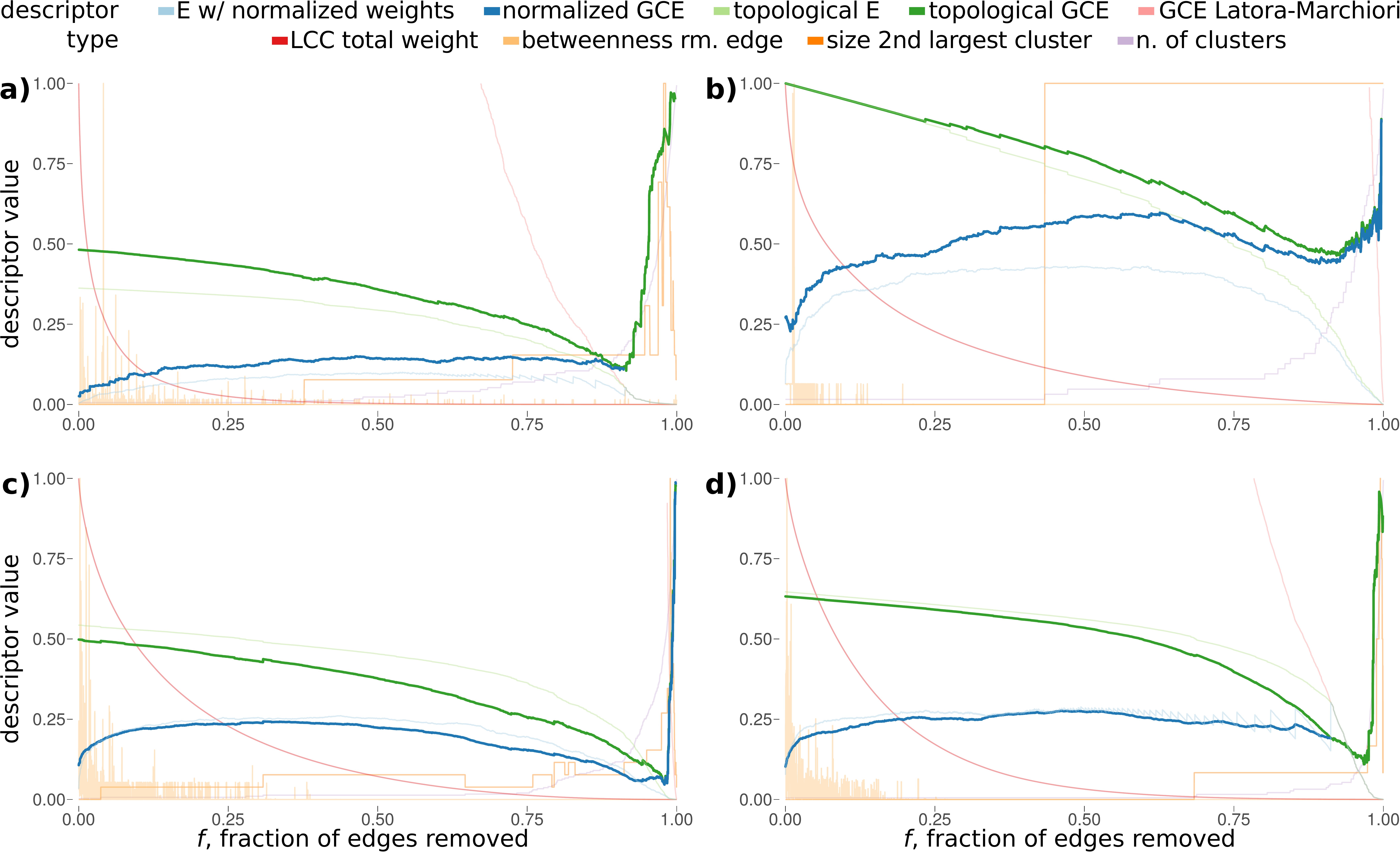}
\caption{\label{fig:real}Targeted bond percolation of real interconnected systems. \textbf{a)} FAO - tea layer, \textbf{b)} Migrations - Vietnam, \textbf{c)} Airports and \textbf{d)} Human brain. Edges are removed in decreasing weight order. Values on the $y-$axis have been cut or rescaled to 1 and represent the quantities described in the legend above.}
\end{figure*}

Additionally, we illustrate how to use the normalization procedure to build a slightly modified version of the GCE that plays the role of a weighted integrity descriptor during bond percolation.
Let $G_0 = G$ and $G_f$ be the network $G$ after the removal of a fraction $f$ of edges and let $E(G_0^{\text{ideal}})$ be the idealized network corresponding to $G_0$ build as described in the Letter.
Then
\begin{equation}
    \text{GCE}^*(G) = \frac{E(G)}{E(G_0^{\text{ideal}})}
\end{equation}
is normalized in $[0, 1]$ and it is a monotone decreasing function w.r.t. $f$.  Figure~\ref{fig:gce-monotone} shows the behaviour of $\text{GCE}^*(G)$ -- both topological and weighted -- for two of the real networks from Tab.1(main Letter).

\begin{figure}[h]
    \includegraphics[width=.45\textwidth]{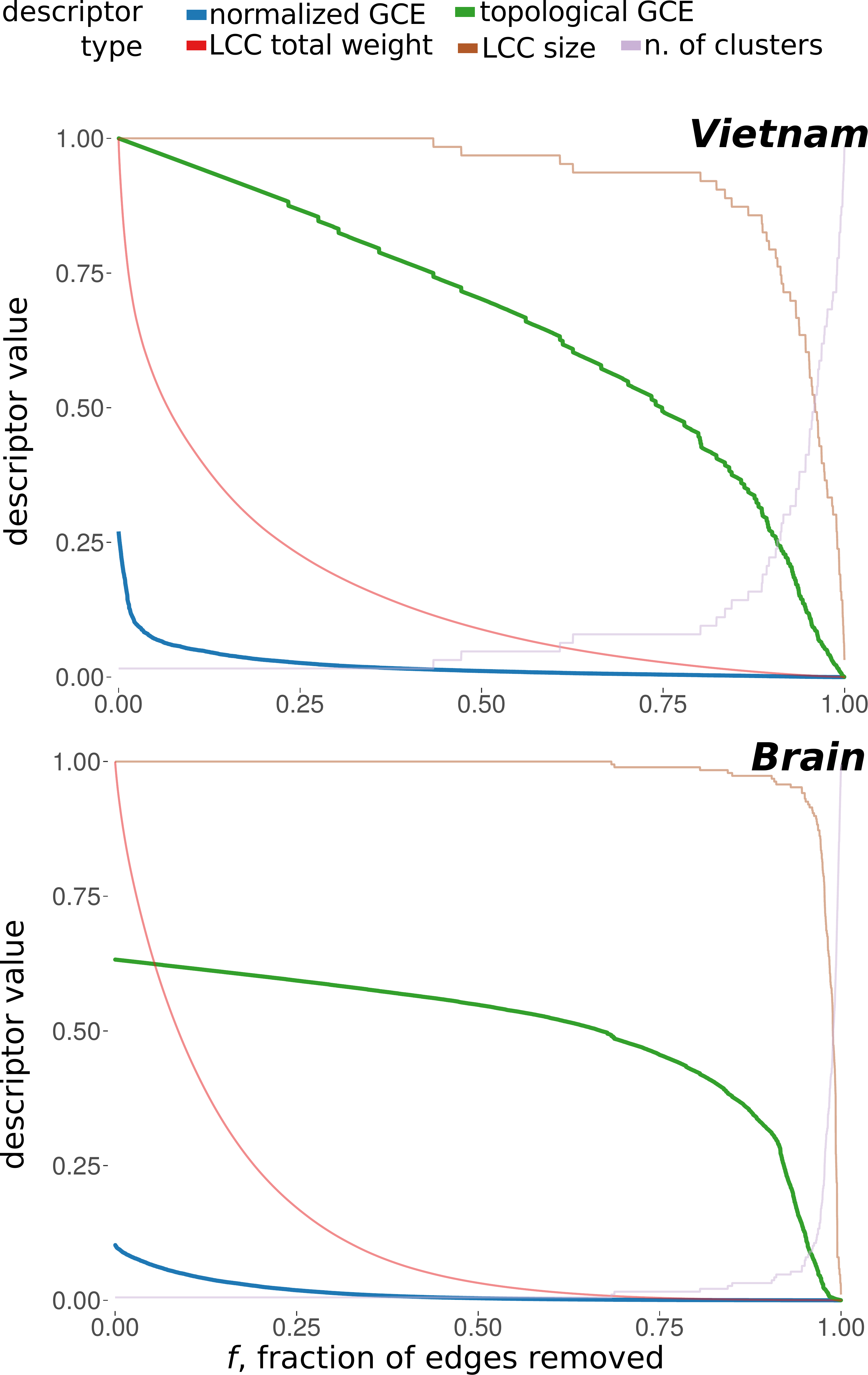}
    \caption{Weighted and topological GCE normalized on the ideal proxy $G_0^{\text{ideal}}$ of the original network, at each step of the percolation simulation.}
    \label{fig:gce-monotone}
\end{figure}

\begin{figure*}[tb]
    \includegraphics[width=.95\textwidth]{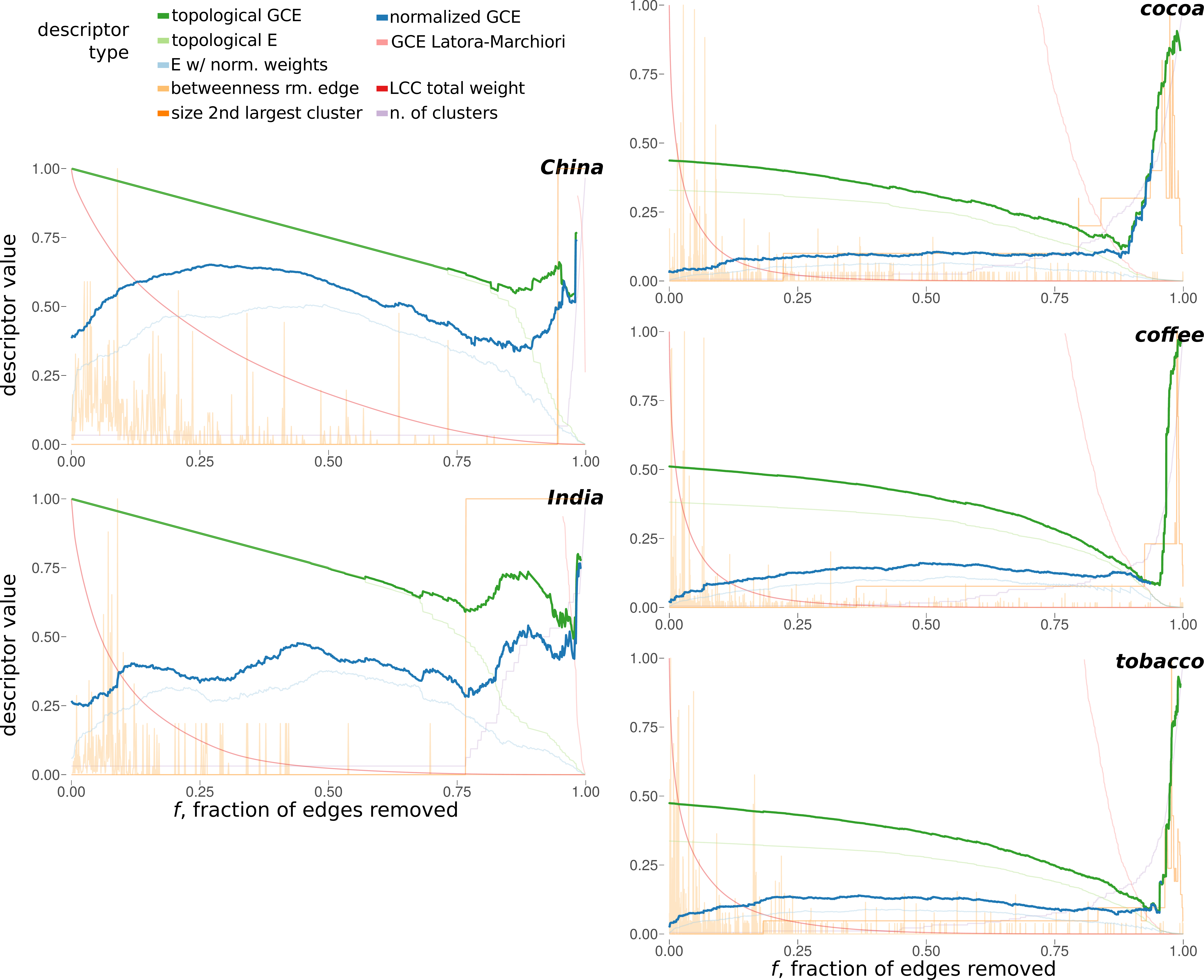}
    \caption{Internal migration (left) and FAO trade network (right) - percolation plots. Values on the $y-$axis have been cut or rescaled to 1 and represent the quantities described in the legend.}
    \label{fig:real-continued}
\end{figure*}

Finally, we show the percolation plots for the remaining datasets studied in this work, see Figures~\ref{fig:real-continued}.

\heading{On the normalized weighted efficiency of Latora and Marchiori \citep{latora2003economic}}
Let us take $G$ as the subgraph consisting of vertices $q, r, v$ of Fig.1 of the main Letter and suppose that the weights are the result of the aggregation of multiple binary connections. Its weighted adjacency matrix is
\begin{equation*}
    \mathbf{W} = \begin{pmatrix}
    \cdot & 4 & 1 \\
    4 & \cdot & 2 \\
    1 & 2 & \cdot
    \end{pmatrix}
\end{equation*}
We can compute physical distances $\ell_{ij}$ following the suggestions in \citep{latora2003economic}
\begin{equation*}
    \mathbf{L} = \begin{pmatrix}
    \cdot & \frac14 & 1 \\
    \frac14 & \cdot & \frac12 \\
    1 & \frac12 & \cdot
    \end{pmatrix}
\end{equation*}
and shortest-path distances minimizing the sum of costs (i.e. inverse weights)
\begin{equation*}
    \mathbf{D} = \begin{pmatrix}
    \cdot & \frac14 & \frac34 \\
    \frac14 & \cdot & \frac12 \\
    \frac34 & \frac12 & \cdot
    \end{pmatrix}
\end{equation*}
The global communication efficiency defined in \citep{latora2003economic} is given by $E_{\mathrm{glob}}=\frac{E(G)}{E\left(G_{\mathrm{ideal}}\right)}$, where $E(G)=\frac{1}{N(N-1)} \sum\limits_{i \neq j} \frac{1}{d_{ij}}$ and $E(G_{\mathrm{ideal}})=\frac{1}{N(N-1)} \sum\limits_{i \neq j} \frac{1}{\ell_{ij}}$.
Observe that the condition (which is sufficient for $\frac{E(G)}{E(G_{\mathrm{ideal}})} \leq 1$)
\begin{equation}\label{eq:dl-condition}
    d_{ij} \geq \ell_{ij} \quad \forall i\neq j \in V_G
\end{equation}
is not satisfied for $i=1, j=3$ and this causes $\text{GCE}=\frac{E(G)}{E(G_{\mathrm{ideal}})}= \frac{22}{9}\left(\frac{7}{3}\right)^{-1} > 1$.

This counter-example on the statement \eqref{eq:dl-condition} is not a pathological case: \eqref{eq:dl-condition} is violated whenever the weighted shortest-path between adjacent nodes $i, j$ does not traverse the direct link $e_{ij}$, i.e. $d_{ij} < \frac{1}{w_{ij}}$ and it may often happen in real networks with large heterogeneous weights.

\begin{figure}[!h]
    \includegraphics[width=.4\textwidth]{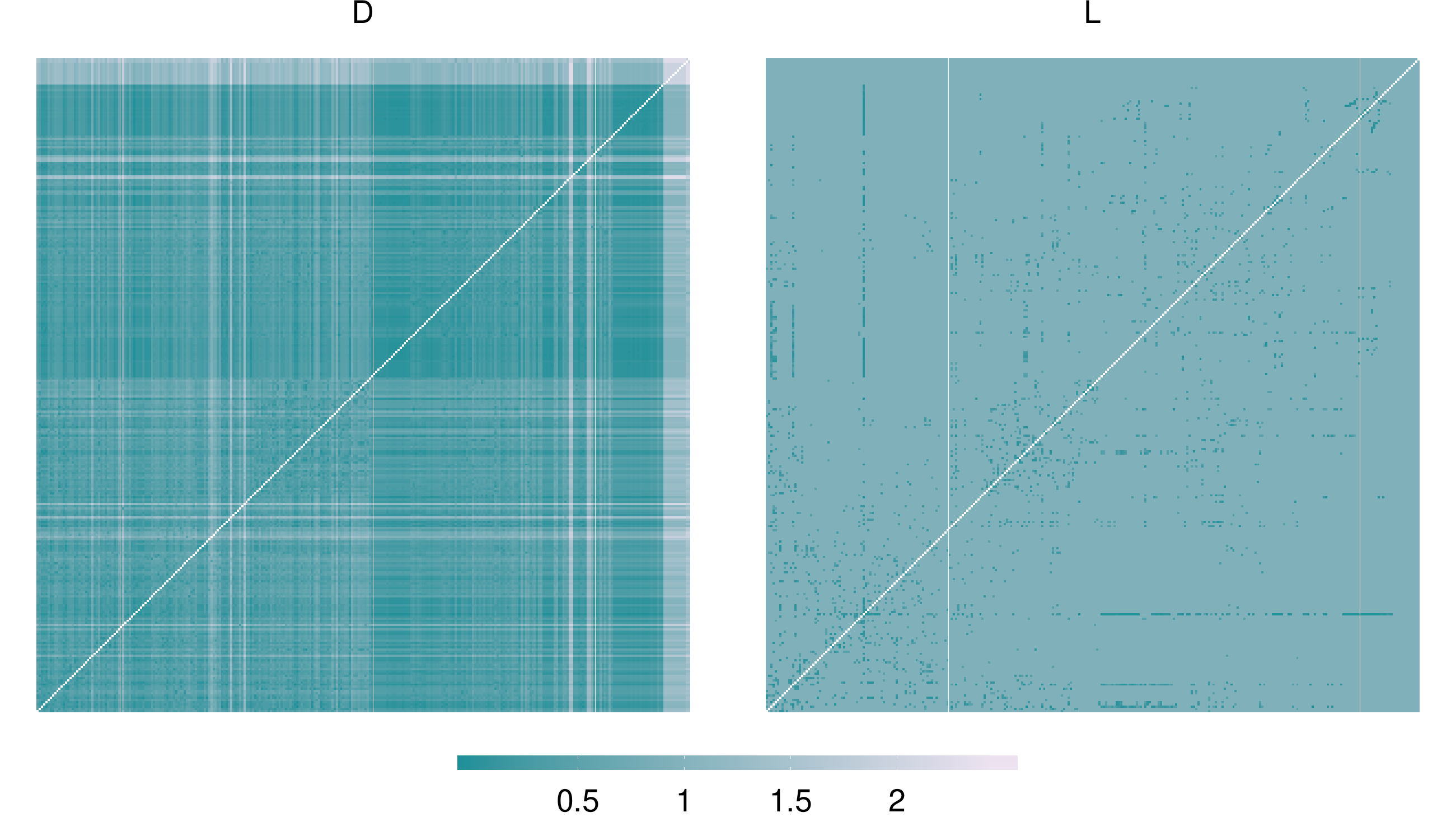}
    \caption{Shortest-path and physical distance matrices of the (undirected) \emph{C. elegans} neural network. 85.6\% of the entries of $\mathbf{D}$ are strictly smaller than their counterparts in $\mathbf{L}$. Some efficiency values $E_{global}=2.52$, topological $E = 0.44$.}
    \label{fig:celeg}
\end{figure}

Trying to reproduce the results in \citep{latora2003economic}, we considered the neural network of the \emph{C. elegans} \citep{watts1998collective,latora2003economic}, with data from \url{http://www-personal.umich.edu/~mejn/netdata/}.
Firstly, we aggregate multiple edges, obtaining a simple, directed, weighted network with $N=297$ nodes, $m = 2345$ edges and weights in the range $[1, 70]$.
If we consider the network as undirected, we obtain $m = 2148$ edges and weights in the range $[1, 72]$. The data are not the same used in \citep{latora2003economic}, so we cannot reproduce their results exactly. Let us focus on the undirected network: Fig.\ref{fig:celeg} shows the distance matrix $D$ evaluated using Dijkstra's algorithm with the reciprocal of edge weights, and the matrix of physical distances $L$, with $\ell_{ij} = \min\{1, \frac{1}{w_{ij}}\}$.

\bibliographystyle{apsrev}